\documentclass[twocolumn,aps,pra,groupedaddress,showpacs]{revtex4}
\usepackage{epsfig}

\begin{document}

\title{Characterizing the entanglement of bipartite 
quantum systems}

\author{Vittorio Giovannetti}
\affiliation{Research Laboratory of Electronics,
MIT - Cambridge, MA 02139, USA}

\author{Stefano Mancini}
\affiliation{INFM, Dipartimento di Fisica, Universit\`a di
Camerino, I-62032 Camerino, Italy}

\author{David Vitali}
\affiliation{INFM, Dipartimento di Fisica, Universit\`a di
Camerino, I-62032 Camerino, Italy}

\author{Paolo Tombesi}
\affiliation{INFM, Dipartimento di Fisica, Universit\`a di
Camerino, I-62032 Camerino, Italy}

\begin{abstract}
We derive a separability criterion for bipartite 
quantum systems which generalizes the already known criteria.
It is based on observables having generic commutation relations.
We then discuss in detail the relation among these criteria.
\end{abstract}

\pacs{03.65.Ud, 03.65.Ta, 03.67.-a}

\maketitle

\section{Introduction}
\label{intro}

Entangled states have been known almost from the very beginning of 
quantum mechanics \cite{SCH,EPR},
and their somewhat unusual features have been 
investigated for many years. However, recent developments in the 
theory of quantum information \cite{QI} 
have required a deeper knowledge of their properties.

The simplest system where one can study entanglement
is represented by a bipartite system.
In such a system, either with discrete or continuous
variable, the inseparability of pure states, 
is now well understood 
and the von Neumann entropy of either subsystem 
quantifies the amount of entanglement \cite{BEN}.
Instead, the question of inseparability of mixed states is 
much more complicated and involves subtle effects.
For discrete variable systems the Peres-Horodecki criterion \cite{PH}
constitutes a theoretical tool to investigate the separability.
Recently, different criteria have been also proposed for continuous 
variable systems \cite{REID,REID1,TAN,DUA,SIM,PRL02}.
Nevertheless a unifying criterion, of practical use, 
does not yet exist.
Needless to say that also the entanglement quantification 
for mixed states is not well assessed \cite{VID}.
On the other hand, the lack of knowledge for bipartite entanglement is 
not only a serious drawback in the study of mixed-state entanglement, 
but also a limitation for understanding multipartite entanglement.

The aim of this paper is to throw some light on the plethora of entanglement
criteria for bipartite systems. In particular, we shall derive a general
separability criterion valid for any state of any
bipartite system. We shall then discuss its relation with the already
known criteria.

\section{A general separability criterion}
\label{section2}

Let us consider a bipartite system whose subsystems, 
not necessarily identical,
are labeled as $1$ and $2$, 
and a separable state ${\hat\rho}_{sep}$
on the Hilbert space  
${\cal H}_{tot}={\cal H}_{1} \otimes {\cal H}_{2}$. 
Such state can be written as
\begin{eqnarray}
    {\hat\rho}_{sep}=\sum_{k} w_{k} \;
    {\hat\rho}_{k1} \otimes {\hat\rho}_{k2}\,,
    \label{sepone} 
\end{eqnarray}
where ${\hat\rho}_{kj}$ ($j=1,2$) are normalized density matrices 
on ${\cal H}_{j}$ while $w_{k}\geq 0$ with
$\sum_{k} w_{k}=1$.

Let us now choose a generic couple of observables for each subsystem,
say  ${\hat r}_{j},{\hat s}_{j}$ on ${\cal H}_{j}$ ($j=1,2$),
and define the operators
\begin{eqnarray}
    {\hat{\cal C}}_{j}= i \left[{\hat r}_{j},{\hat s}_{j} 
    \right]\,,
    \quad j=1,2\,.
    \label{commmm}
\end{eqnarray}
Notice that the two couples ${\hat r}_{j},{\hat s}_{j}$ 
may represent completely different observables, e.g.
one couple may refer to a continuous variable subsystem
while the other to a discrete variable subsystem.
Furthermore, ${\hat{\cal C}}_{j}$ is typically nontrivial Hermitian
operator on the Hilbert subspaces. 

We now introduce the following observables on ${\cal H}_{tot}$:
\begin{eqnarray}
    {\hat u} &=& a_{1} {\hat r}_{1} + a_{2} {\hat r}_{2} \,,
    \nonumber \\
    {\hat v} &=& b_{1} {\hat s}_{1} + b_{2} {\hat s}_{2} \,,
    \label{unouno}
\end{eqnarray}
where $a_{j},b_{j}$ are real parameters.
From the standard form of the uncertainty principle \cite{SAK}, 
it follows that
every state ${\hat\rho}$ on ${\cal H}_{tot}$ must satisfy 
\begin{eqnarray}
    \langle (\Delta {\hat u})^{2} \rangle \langle 
    (\Delta {\hat v})^{2} \rangle 
    \geq \frac{| a_{1} b_{1} \langle {\hat{\cal C}}_{1}  \rangle + 
    a_{2} b_{2} \langle {\hat{\cal C}}_{2}  \rangle|^{2} }{4}\,,
\label{treone} 
\end{eqnarray}
where $\langle \hat{\Theta} \rangle \equiv {\rm Tr}[ \hat{\Theta} {\hat\rho} ]$
is the expectation value over ${\hat\rho}$ of the operator 
$\hat{\Theta}$, and $\Delta \hat{\Theta} \equiv \hat{\Theta} - \langle 
\hat{\Theta}\rangle$.
However, for separable states, a stronger bound exists.
As a matter of fact, the following theorem holds

\vspace{1cm}

{\bf Theorem}:
{\it 
\begin{eqnarray}
    {\hat\rho}_{sep}\quad\Longrightarrow\quad
    \langle (\Delta {\hat u})^{2} \rangle 
    \langle (\Delta {\hat v})^{2} \rangle 
    \geq {\tilde{\cal O}^{2}}\,,
\label{newsep} 
\end{eqnarray}
with
\begin{eqnarray}
   \tilde{\cal O}= \frac{1}{2} \big( \;| a_{1} b_{1}| \;  \tilde{\cal O}_{1}  
   + |a_{2} b_{2}| \; \tilde{\cal O}_{2} \; \big)\,,
\label{atilde} 
\end{eqnarray}
where
\begin{eqnarray}
    \tilde{\cal O}_{j} \equiv \sum_{k} w_{k} \; 
    |\langle {\hat{\cal C}}_{j} \rangle_{k}|\,, \qquad j=1,2,
    \label{ajtilde} 
\end{eqnarray}
being $\langle {\hat{\Theta}}_{j} \rangle_{k} \equiv {\rm Tr}
[  {\hat{\Theta}}_{j}  {\hat\rho}_{kj} ]$
the expectation value of the operator ${\hat{\Theta}}_{j}$
onto $\hat{\rho}_{kj}$}.

\vspace{1cm}

{\bf Proof}:
From the definitions of $\langle (\Delta {\hat u})^{2} \rangle$,
and ${\hat\rho}_{sep}$ it is easy to see that
\begin{eqnarray}
    \langle (\Delta {\hat u})^{2} \rangle &=& 
    \sum_{k} w_{k} \left[ \; a_{1}^{2} \;
    \langle (\Delta {\hat r}_{1}^{(k)})^{2} \rangle_{k}
    + a_{2}^{2} \;\langle (\Delta {\hat r}_{2}^{(k)})^{2} \rangle_{k}
    \;\right]   
    \nonumber
    \\ &&+ \sum_{k} w_{k} \langle {\hat u} \rangle_{k}^{2} 
    - \left(\sum_{k} w_{k} 
    \langle {\hat u} \rangle_{k}\right)^{2} 
    \label{tt1}\,,
\end{eqnarray}
where the quantity 
$\Delta {\hat r}_{j}^{(k)}\equiv
{\hat r}_{j}-\langle {\hat r}_{j} \rangle_{k} $ is 
the variance of the operator $\hat{r}_{j}$ on the state $\hat{\rho}_{kj}$.
An analogous expression holds for $\langle (\Delta {\hat v})^{2} \rangle$.
By applying the Cauchy-Schwartz inequality on the last two terms of 
the r.h.s. of (\ref{tt1}) we obtain 
\begin{eqnarray}
    \langle (\Delta {\hat u})^{2} \rangle &\geq&  \sum_{k} w_{k} 
    \left[ \; a_{1}^{2} \;
    \langle (\Delta {\hat r}_{1}^{(k)})^{2} \rangle_{k}
    + a_{2}^{2} \;\langle (\Delta {\hat r}_{2}^{(k)})^{2} \rangle_{k}
    \;\right]  \,,
    \nonumber\\ 
   \label{ttt1} 
\end{eqnarray}
and analogously   
\begin{eqnarray}
   \langle (\Delta {\hat v})^{2} \rangle &\geq&
     \sum_{k} w_{k} \left[ \; b_{1}^{2} \;
    \langle (\Delta {\hat s}_{1}^{(k)})^{2} \rangle_{k}
    + b_{2}^{2} \;\langle (\Delta {\hat s}_{2}^{(k)})^{2} \rangle_{k}
    \;\right]  \,.
    \nonumber\\ 
    \label{ttt2}
\end{eqnarray}
Now, taking any two real nonnegative numbers $\alpha$ and 
$\beta$, and using the relations (\ref{ttt1}) and (\ref{ttt2}),
we get the following inequality
\begin{eqnarray}
    &&\alpha \langle (\Delta {\hat u})^{2} \rangle 
    + \beta  \langle (\Delta {\hat v})^{2} \rangle 
    \geq  \label{imp} \\
    &&\sum_{k} w_{k} \; \left[ \alpha \; a_{1}^{2} 
    \langle (\Delta {\hat r}_{1}^{(k)})^{2} \rangle_{k} + 
    \beta \; b_{1}^{2} 
    \langle (\Delta {\hat s}_{1}^{(k)})^{2} \rangle_{k} \right] 
    \nonumber\\
    &&+ \sum_{k} w_{k} \; \left[ \alpha \;  a_{2}^{2} 
    \langle (\Delta {\hat r}_{2}^{(k)})^{2} \rangle_{k} + 
    \beta \; b_{2}^{2} 
    \langle (\Delta {\hat s}_{2}^{(k)})^{2} \rangle_{k} \right]\,.
    \nonumber
\end{eqnarray}
Furthermore, by applying the uncertainty principle 
to the operators $\hat{r}_{j}$ and $\hat{s}_{j}$ on the state $\hat{\rho}_{kj}$,
it follows
\begin{eqnarray}
    &&\alpha \; a_{j}^{2} 
    \langle (\Delta {\hat r}_{j}^{(k)})^{2} \rangle_{k} + 
    \beta \; b_{j}^{2} 
    \langle (\Delta {\hat s}_{j}^{(k)})^{2} \rangle_{k} \geq
    \nonumber\\
    &&\alpha \; a_{j}^{2} 
    \langle (\Delta {\hat r}_{j}^{(k)})^{2} \rangle_{k} + 
    \beta \; b_{j}^{2}\frac{|\langle {\hat{\cal C}}_{j} 
    \rangle_{k}|^{2}}{4\langle (\Delta {\hat r}_{j}^{(k)})^{2} \rangle_{k}}
    \geq 
    \nonumber\\
    &&\sqrt{\alpha \beta} \; | a_{j} b_{j} | \; |\langle 
    {\hat{\cal C}}_{j}\rangle_{k} |\,.
    \label{princcc}
\end{eqnarray}

The last inequality of Eq.(\ref{princcc}) 
comes from the behavior, for $x\geq 0$,
of the function 
\begin{eqnarray}
    f(x)= \gamma_{1} x + \gamma_{2}/x
    \label{effe}\,, 
\end{eqnarray}
with $\gamma_{1}, \gamma_{2}\geq 0$.
Such function takes the minimum value 
$f_{min}= 2 \sqrt{ \gamma_{1} \gamma_{2} }$.
Then, inserting Eq. (\ref{princcc}) into (\ref{imp}) we obtain
\begin{eqnarray}
    \alpha \langle (\Delta {\hat u})^{2} \rangle 
    +\beta  \langle (\Delta {\hat v})^{2} \rangle 
    &\geq& 2 \sqrt{\alpha \beta} \; \tilde{\cal O}
    \label{veryimp}\,.
\end{eqnarray}
where $\tilde{\cal O}$ has been defined in Eq.(\ref{atilde}).

Notice that, for a given system state, the quantities  
$\langle (\Delta {\hat u})^{2} \rangle$, 
$\langle (\Delta {\hat v})^{2} \rangle$ 
and $\tilde{\cal O}$ are fixed, and the inequality
(\ref{veryimp}) must be satisfied for every positive value
of $\alpha$ and $\beta$. Thus, we can write 
\begin{eqnarray}
    \langle (\Delta {\hat u})^{2} \rangle 
    &\geq& \max_{\alpha > 0; \; \beta\geq 0} \left\{ 2 
    \sqrt{\frac{\beta}{\alpha}} \; \tilde{\cal O} - \frac{\beta}{\alpha}  
    \langle (\Delta {\hat v})^{2} \rangle 
    \right\}
    \nonumber\\ 
    &=& \frac{\tilde{\cal O}^{2}}{\langle (\Delta v)^{2} 
    \rangle}\,,
    \label{veryimpimp}
\end{eqnarray}
where the equality has been obtained by maximizing
the function
$g(x)= 2 \, x \, \tilde{\cal O} - x^{2}  \langle (\Delta {\hat v})^{2} \rangle$
over $x\geq 0$.
This concludes the proof of Eq.(\ref{newsep}). 

\begin{figure}[t]
\begin{center}
\epsfxsize=.80\hsize\leavevmode\epsffile{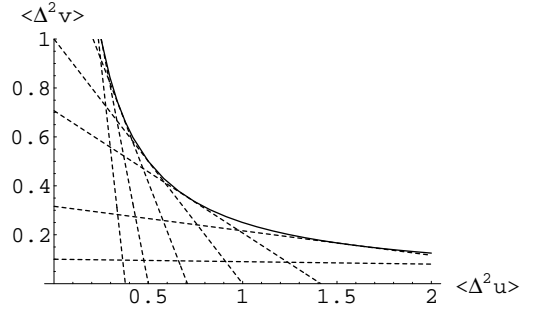}
\end{center}
\caption{Each dashed line represents the boundary defined
by Eq.(\ref{veryimp}) for a given $\alpha$ and $\beta$. 
The envelope of such lines determines
the hyperbola corresponding to Eq.(\ref{newsep}). 
The parameters in the plot are in arbitrary units.}
\label{fig1}
\end{figure}

Practically, proving the theorem,
we have created a family of linear inequalities
(\ref{veryimp}), which must be always satisfied by separable
states. The ``convolution" of such relations gives
the condition (\ref{newsep}), representable by a 
region in the $\langle (\Delta\hat{u})^{2} \rangle$,
$\langle (\Delta\hat{v})^{2} \rangle$ plane delimited by an
hyperbola (see Fig.\ref{fig1}). 
Notice also that, since $\tilde{{\cal O}}_{j} =
\sum_{k} w_{k} \; |\langle {\hat{\cal C}}_{j} 
\rangle_{k}|\geq |\sum_{k} w_{k} \; \langle {\hat{\cal C}}_{j} 
\rangle_{k}|= | \langle {\hat{\cal C}}_{j} \rangle|$, 
the following inequalities hold
\begin{eqnarray}
    \tilde{\cal O}
    &\geq& \frac{1}{2} \big( \;|a_{1}b_{1}|\;|\langle{\hat{\cal C}}_{1}\rangle|
    +|a_{2}b_{2}|\;|\langle{\hat{\cal C}}_{2}\rangle| \; \big)
    \label{dsatilde2}\\    
    &\geq& \frac{1}{2} \big( \; | a_{1} b_{1}\;\langle{\hat{\cal C}}_{1}\rangle +
    a_{2} b_{2}\;\langle{\hat{\cal C}}_{2}\rangle | \; \big)\,.
    \label{dsatilde3} 
\end{eqnarray}
In particular, Eq. (\ref{dsatilde3}) 
tells us that the bound (\ref{newsep}) 
for separable states is much stronger than 
Eq. (\ref{treone}) for generic states.
Moreover, Eq. (\ref{dsatilde2}) gives us a simple separability criterion.
In fact, while $\tilde{\cal O}$ is not easy to evaluate directly, 
as it depends on the type of convex decomposition
(\ref{sepone}) one is considering,
the right hand side of Eq. (\ref{dsatilde2}) is 
easily measurable, as it depends on the
expectation value of the 
observables ${\hat{\cal C}}_{j}$. 
In this sense we can claim that Eq.(\ref{newsep}) 
is a necessary criterion for separability, i.e. 
\begin{eqnarray}
    \langle (\Delta {\hat u})^{2} \rangle 
    \langle (\Delta {\hat v})^{2} \rangle 
    < \tilde{\cal O}^{2}
    \quad\Longrightarrow\quad
    {\hat\rho}\quad{\rm entangled}\,.
\label{newent} 
\end{eqnarray}
An important simplification applies when 
the observable $\hat{\cal C}_{j}$ 
is proportional to the identity operator (e.g. 
 $\hat{r}_{j}$ is the position and $\hat{s}_{j}$ is
 the momentum operator of a particle),
or more generally when it is positive (or negative) definite.
In this case the inequality (\ref{dsatilde2}) 
reduces to an identity and the quantity $\tilde{\cal O}$ does not
depends on the convex decomposition (\ref{sepone}).

The criterion (\ref{newsep}) can be further generalized 
if one adopts the strong version of the uncertainty principle
\cite{SAK} in deriving the inequality (\ref{princcc}).
In this situation the quantity $\tilde{\cal O}_{j}$ of Eq. (\ref{atilde} )
becomes,
\begin{eqnarray}
    \tilde{\cal O}_{j} \equiv 2 \sum_{k} w_{k} \; 
    |\langle \Delta\hat{r}_{j}^{(k)} \Delta\hat{s}_{j}^{(k)} 
    \rangle_{k}|\,, \qquad j=1,2,
    \label{ajtilde1} 
\end{eqnarray}
where $ \Delta\hat{r}_{j}^{(k)}$ and $\Delta\hat{s}_{j}^{(k)} $
are the same objects we have introduced in Eqs. (\ref{ttt1}) and
(\ref{ttt2}). Also in this case $\tilde{\cal O}$ depends
in general on the convex decomposition (\ref{sepone}) of the 
state $\hat{\rho}$.

\section{Relation with other criteria}

In this Section we analyze the relation between the
criterion (\ref{newsep}) and other necessary criteria for
separability that have been proposed in the past. 

First of all it is possible to show that the ``sum" criterion 
of Ref. \cite{DUA} represents a particular case of Eq. (\ref{newsep}).
As a matter of fact the ``sum" criterion is given by
Eq. (\ref{veryimp}) with $\alpha=\beta=1$, 
\begin{eqnarray}
    \langle (\Delta\hat{u})^{2} \rangle 
    + \langle (\Delta\hat{v})^{2} \rangle \;
    &\geq&  \; 2 \; \tilde{\cal O}
    \nonumber\\ 
    &\geq&  \; |a_{1} b_{1}| |  \langle \hat{\cal C}_{1}  
    \rangle | + 
    |a_{2} b_{2}| \; | \langle \hat{\cal C}_{2}  \rangle |
    \label{DUAN} \; ,\nonumber\\
\end{eqnarray}
where we have exploited Eq. (\ref{dsatilde2})
to get a rhs independent from the convex decomposition of 
$\hat{\rho}_{sep}$.
The fact that the ``sum" criterion
comes from condition (\ref{newsep})
is a consequence of the fact that the latter 
has been derived by maximizing over the family of inequalities
(\ref{veryimp}) [see Eq. 
(\ref{veryimpimp}) and Fig. (\ref{fig1})].
However, a straightforward derivation is easy to obtain as well.
In fact, from Eq.(\ref{newsep}) we have
\begin{eqnarray}
    \langle (\Delta\hat{u})^{2} \rangle 
    \; + \;\langle (\Delta\hat{v})^{2} \rangle \;
    \geq   \langle (\Delta\hat{u})^{2} \rangle \; + \; 
    \frac{\tilde{\cal O}^{2}}{\langle (\Delta\hat{u})^{2} \rangle} 
    \; \geq \;  2\tilde{\cal O}\; ,\nonumber\\
    \label{DUANduan} 
\end{eqnarray}
where, for the second inequality we have used
the property of $f(x)$ in Eq. (\ref{effe}). 

Let us now compare the criterion developed in Section~\ref{section2}
with the ``product" criterion developed in Ref. \cite{PRL02}.
The latter, with the generic operators ${\hat u}$ 
and ${\hat v}$ of Eq. (\ref{unouno}),
can be written as
\begin{eqnarray}
    \langle (\Delta\hat{u})^{2} \rangle 
    \langle (\Delta\hat{v})^{2} \rangle \;
    \geq \; |a_{1} a_{2} b_{1} b_{2}| \;
    \frac{| \langle {\hat{\cal C}}_{1} 
    \otimes {\hat{\cal C}}_{2} \rangle |^{2} }{|| 
    {\hat{\cal C}}_{1}|| \, || {\hat{\cal C}}_{2}||}
\label{treTTTttt} \; ,
\end{eqnarray}
where
\begin{eqnarray}
    || {\hat{\cal C}}_{j}||\; = 
    \;\sup_{|\psi\rangle \in {\cal H}_{j}} \;
    \left\{ \; \left|\langle \psi | 
    {\hat{\cal C}}_{j} | \psi 
    \rangle \right|  \;  \right\}
    \label{quattroqqqq} \; .
\end{eqnarray}
is the norm of the operator ${\hat{\cal C}}_{j}$ ($j=1,2$).
In order to proof that Eq.(\ref{treTTTttt}) comes from 
Eq.(\ref{newsep}), we first note that the properties of the
function $f(x)$ of Eq. (\ref{effe}) allows us to write
\begin{eqnarray}
   \tilde{{\cal O}} &=& \frac{\sqrt{| a_{1} a_{2} b_{1} b_{2}|}}{2} \;  
    \left[ \; \sqrt{\frac{| a_{1} b_{1}|}{| a_{2} b_{2}|}}  \;
   \tilde{{\cal O}_{1}} \; + \;
   \sqrt{\frac{| a_{2} b_{2}|}{| a_{1} b_{1}|}}  \;
   \tilde{{\cal O}_{2}} \right]
    \nonumber\\
    &\geq&\;\sqrt{| a_{1} a_{2} b_{1} b_{2}|}  \;
    \sqrt{ \;\tilde{{\cal O}_{1}}\tilde{{\cal O}_{2}}  \;}
\label{aaatilde} \; .
\end{eqnarray}

Then, by applying the Cauchy-Schwarz inequality and 
using the definitions (\ref{ajtilde}) and (\ref{quattroqqqq})
we can build up the following chain of relations
\begin{eqnarray}
    &&| \langle {\hat{\cal C}}_{1} \otimes 
    {\hat{\cal C}}_{2} \rangle |^{2} 
    \; \equiv \;
    \left| \; \sum_{k} w_{k} \; \langle {\hat{\cal C}}_{1} \rangle_{k} 
    \; \langle {\hat{\cal C}}_{2} 
    \rangle_{k} \;  \right|^{2}
    \nonumber\\ 
    &&\leq  
    \left[ \sum_{k} w_{k} \; |\langle \hat{\cal C}_{1} 
    \rangle_{k}|^{2} \right] \; \left[ \sum_{k} w_{k} \; |
    \langle \hat{\cal C}_{2} 
    \rangle_{k}|^{2} \right] \nonumber \\
    &&\leq  \; || \hat{\cal C}_{1} || \; || \hat{\cal C}_{2} || \; 
    \left[ \sum_{k} w_{k} \; |\langle \hat{\cal C}_{1} 
    \rangle_{k}|  \right] \; \left[ \sum_{k} w_{k} \; |\langle 
    \hat{\cal C}_{2} 
    \rangle_{k}|  \right] 
    \nonumber\\
    &&\equiv || \hat{\cal C}_{1} || \; || \hat{\cal C}_{2} || \;
    \tilde{\cal O}_{1} \; \tilde{\cal O}_{2} 
    \label{diseg} \; ,
\end{eqnarray}
or
\begin{eqnarray}
    \tilde{\cal O}_{1} \; \tilde{\cal O}_{2} 
    \geq \frac{| \langle \hat{\cal C}_{1} 
    \otimes \hat{\cal C}_{2} \rangle |^{2}}
    {|| \hat{\cal C}_{1} || \; || \hat{\cal C}_{2} ||}
\label{relazione} \; .
\end{eqnarray}
Substituting  Eqs. (\ref{aaatilde}) and (\ref{relazione}) into 
Eq. (\ref{newsep}) we finally get Eq. (\ref{treTTTttt}).
Equations (\ref{newsep}) and (\ref{treTTTttt}) give the same
separability criterion when $\hat{\cal C}_{j}$ is a
real number $c_{j}$, and the 
parameters $a_{j}, b_{j}$ satisfy the condition 
$a_{1} b_{1} c_{1} =\pm a_{2} b_{2} c_{2}$.
An example of this situation has been presented in \cite{PRL02}.

\begin{figure}[t]
\begin{center}
\epsfxsize=.80\hsize\leavevmode\epsffile{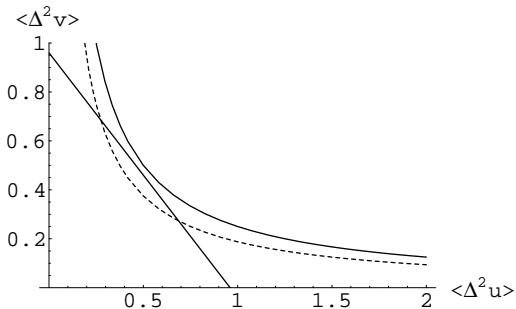}
\end{center}
\caption{Partition of the plane
$\langle (\Delta\hat{u})^{2} \rangle$, 
$\langle (\Delta\hat{v})^{2} \rangle$
based on different 
state separability conditions.
See the text for details.}
\label{fig2}
\end{figure}

Summarizing, we have proved that condition
(\ref{newsep}) is stronger than the criteria of 
Refs. \cite{DUA,PRL02}.
This is depicted in Fig. \ref{fig2}, where the inequality
(\ref{newsep}) determines a zone under the 
solid hyperbola where we can 
only find entangled states: separable states 
must lie above this curve. 
Notice however that entangled states could also lie above the solid
hyperbola since the condition (\ref{newsep}) is only sufficient for
entanglement.
On the other hand, also the condition (\ref{treTTTttt})
determines a portion of the plane  
where only entangled states can live:
that below the dashed hyperbola.
However, this part is entirely included in the portion 
subtended by the solid hyperbola. 
Finally, criterion (\ref{DUAN}) determines 
a straight line inclined at $-45^{\circ}$ which in general is not
tangent to the solid hyperbola representing condition
(\ref{newsep}). 
Also in this case the portion of the 
plane reserved to an entangled 
states is included
in the portion delimited by the hyperbola of Eq. (\ref{newsep}).
This shows the generality of the criterion presented in 
Section~\ref{section2}.

A couple of interesting connections can be also established 
when comparing the
criterion of Eq. (\ref{newsep}) with the  {\em weaker EPR criterion} 
discussed in \cite{REID1} and with Simon criterion \cite{SIM}. 
In fact Eq. (\ref{newsep}) and the weaker EPR 
criterion are essentially equivalent when 
applied to observables $\hat{r}_{j}$, $\hat{s}_{j}$
with trivial commutation rules.
In order to show this, it is sufficient to observe
that the uncertainties $\langle (\Delta {\hat u})^{2} \rangle $ and 
$\langle (\Delta {\hat v})^{2} \rangle $ give 
an upper bound for the errors in the inferred measurements
 of the observables $a_{1} \hat{r}_{1}$ and $b_{1} \hat{s}_{1}$ 
 obtained through a direct measurement of the operators 
$-a_{2}\hat{r}_{2}$ and $-b_{2}\hat{s}_{2}$ on $\hat{\rho}$ 
(see \cite{REID,REID1} for more details about 
the definition of the inferred measurements).
The comparison 
with Simon's criterion is obtained 
considering the case in which $\hat{r}_{j}$, $\hat{s}_{j}$
are linear combinations of the position $\hat{q}_{j}$ and momentum
$\hat{p}_{j}$ operators of the $j-$th system, i.e.
\begin{eqnarray}  
   \hat{r}_{1} \equiv \hat{q}_{1} + \frac{a_{3}}{a_{1}} 
   \hat{p}_{1} &\quad \quad&
   \hat{s}_{1} \equiv \hat{p}_{1} + \frac{b_{3}}{b_{1}} 
   \hat{q}_{1} \nonumber \\
   \hat{r}_{2} \equiv \hat{q}_{2} + \frac{a_{4}}{a_{2}} 
   \hat{p}_{2} &\quad \quad&
   \hat{s}_{2} \equiv \hat{p}_{2} + \frac{b_{4}}{b_{2}} 
   \hat{q}_{2}
\label{newqp} \; , 
\end{eqnarray}
where $a_{3}, a_{4}, b_{3}$ and $b_{4}$ are generic real parameters. 
Since in this case $\left[\hat{q}_{j},\hat{p}_{j}\right]=i$, 
Eq. (\ref{newsep}) becomes
\begin{eqnarray}
    \langle (\Delta u)^{2} \rangle \langle (\Delta v)^{2} \rangle 
    \geq \frac{1}{4}\big(\;| a_{1} b_{1} -a_{3} b_{3}| 
    +  |a_{2} b_{2} - a_{4} b_{4}|\;\big)^{2}, \nonumber \\
\label{nostro1} 
\end{eqnarray}
that has to be compared with the corresponding equation of \cite{SIM},
i.e. 
\begin{eqnarray}
\langle (\Delta u)^{2} \rangle + \langle (\Delta v)^{2} \rangle 
    \geq  | a_{1} b_{1} -a_{3} b_{3}|
    +  |a_{2} b_{2} - a_{4} b_{4}| 
\label{simon} \; .
\end{eqnarray}
It is easy to verify that given $a_{j}, b_{j}$ ($j=1, \cdots ,4$), the 
``product" condition of Eq. (\ref{nostro1}) implies 
the ``sum" condition of Eq. (\ref{simon}).
However, the necessary criterion for separability of Simon
requires that Eq. (\ref{simon}) should be verified
for {\em all} possible values of the coefficients $a_{j}, b_{j}$ 
(see Eq. (11) of \cite{SIM}). 
In this case, Eqs. (\ref{nostro1}) and (\ref{simon}) are equivalent 
since
one can reobtain the first from the second using the same
convolution trick already used in deriving (\ref{veryimpimp}) 
from Eq. (\ref{veryimp}). In particular this means that 
Eq. (\ref{nostro1}),
when considered for all possible values of $a_{j}, b_{j}$,  
provides a criterion 
for separability which is necessary {\em and} sufficient 
if applied to Gaussian states.

Finally, it is also possible to establish a connection with the 
criteria used in Refs. \cite{KOR,BOW} for discrete variable systems.
In particular, assuming $\alpha=\beta=1$ in Eq. (\ref{veryimp})
and $a_{1}=1$, $a_{2}=\pm a_{1}$, $b_{1}=1$, $b_{2}=\pm b_{1}$,
in Eq.(\ref{dsatilde2}), we get 
\begin{equation}
\langle(\Delta\hat{u})^{2}\rangle
\langle(\Delta\hat{v})^{2}\rangle\ge
|\langle\hat{\cal C}_{1}\rangle|
+|\langle\hat{\cal C}_{2}\rangle|\,.
\end{equation}
For symmetric condition between the two subsystems, the above equation
reduces to Eq.(4) of Ref. \cite{BOW}
with ${\hat r}_{j}$, ${\hat s}_{j}$ the 
fluctuations of the Stokes parameters.

\section{Conclusion}

In this paper we have studied the connections between the
separability condition of the initial state of a bipartite system
and the uncertainty relation of a  couple of non-local observables 
$\hat{u}$, $\hat{v}$ of the two subsystems. 
In the case where $\hat{u}$, $\hat{v}$
are linear combinations of generic operators $\hat{r}_{j}$, $\hat{s}_{j}$
of the two subsystems, we 
have derived a mathematical constraint, Eq. (\ref{newsep}),
that has to be satisfied by a separable system.
In general this relation depends 
on terms which are not measurable, meaning that 
Eq. (\ref{newsep}) can not be directly used to 
test experimentally the separability of the system. 
However, in many cases of experimental relevance
Eq. (\ref{newsep}) can be expressed in terms of  
measurable quantities (see the discussion at the end of 
Section~\ref{section2}), 
providing a very general necessary criterion for
separability, i.e. a sufficient criterion for entanglement.
Most importantly, Eq. (\ref{newsep}) represents a powerful
theoretical tool which can be used to derive new measurable criteria
(see for instance Eqs. (\ref{dsatilde2}) and (\ref{treTTTttt})) and
to compare them with other already known (e.g. those given 
in Refs. \cite{REID,REID1,PRL02,SIM,DUA,TAN,KOR,BOW}).

\end{document}